\documentclass[showpacs,showkeys,preprint,preprintnumbers,amsmath,amssymb]{revtex4}
\usepackage{graphicx,psfrag}
\usepackage{amsfonts}
 \usepackage{amsmath}
 \usepackage{bm}% bold math
\usepackage{epsfig,subfigure}
\usepackage{color}
\usepackage{CJK}

\def \be {\begin{equation}}
\def \ee {\end{equation}}

\begin{document}
\begin{CJK*}{GB}{gbsn}
%\CJKindent
\preprint{}

\title{Acceleration of particles in Einstein-Maxwell-dilaton black holes}

\author{Pu-Jian Mao$^1$$^,$$^3$}
\author{Ran Li$^2$}\thanks{liran@htu.edu.cn}
\author{Lin-Yu Jia$^3$}
\author{Ji-Rong Ren$^3$}
\thanks{renjr@lzu.edu.cn}
\affiliation{$^1$Institute of High Energy Physics and Theoretical Physics Center for Science Facilities,
Chinese Academy of Sciences, 19B Yuquan Road, Beijing 100049, P. R. China\\
$^2$Department of Physics, Henan Normal University, Xinxiang 453007, China\\
$^3$Institute of Theoretical Physics, Lanzhou University, Lanzhou 730000, China}

\begin{abstract}
It has recently been  pointed out that, under certain conditions, the
energy of particles accelerated by black holes in the center-of-mass
frame can become arbitrarily high. In this paper, we study the
collision of two particles in the case of four-dimensional charged nonrotating, extremal charged rotating and near-extremal charged rotating
Kaluza-Klein black holes as well as the naked singularity case in Einstein-Maxwell-dilaton theory. We find that the
center-of-mass energy for a pair of colliding particles is unlimited
at the horizon of charged nonrotating Kaluza-Klein black holes, extremal charged rotating
Kaluza-Klein black holes and in the naked singularity case.
\end{abstract}

\pacs{97.60.Lf, 04.70.-s, 04.50.Cd }

\keywords{ Black Hole, Particles Accelerator, Center-of-Mass Energy}

\maketitle

\section{Introduction}
The Planck scale defines the meeting point of gravity and quantum
mechanics. The probe of the Planck-scale physics also contributes to
the discovery of extra dimensions of space-time and the Grand
Unification Theory. However, compared with the Planck energy of
$10^{16}$ TeV, the largest terrestrial accelerator, the Large Hadron
Collider, which can detect physics at collision energies of order $10^1$ TeV,
is too low to probe Planck-scale physics. There is a very, very
large gap between the Planck scale and our current experimental
techniques. So some other new physics mechanisms should be proposed for
probing Planck-scale physics. The collision of particles around a
black hole may provide a possible detection.

Ba\~{n}ados, Silk and West (BSW) \cite{1} recently investigated the
maximum center-of-mass energy of particles colliding around a Kerr
black hole. Their result showed that the maximum energy grows with
$a$, which is the unit angular momentum of the black hole.
Remarkably, as the black hole becomes extremal, they found a
fascinating and important property of the extremal Kerr black hole,
that two particles freely falling from rest at spatial infinity can
collide at the horizon with arbitrarily high center-of-mass (CM)
energy. Such a black hole could serve as a particle accelerator and
may provide a visible probe of Planck-scale physics. To achieve that, the black
hole must have a maximized angular momentum and one of the colliding
particles should have orbital angular momentum per unit rest mass
$l=2$. Subsequently, in \cite{2,3}, the authors argued that the CM
energy is in fact limited, because there is always a small
deviation of the spin of an astrophysical black hole from its
maximal value. According to the work of Thorne \cite{Thorne}, the
dimensionless spin of astrophysical black holes should not exceed
$a=0.998$. In terms of the small parameter $\epsilon=1-a$, Jacobson
and Sotiriou got the maximal CM energy
$\frac{E^{max}_{CM}}{m_0}\sim4.06\epsilon^{-1/4}+\mathcal
{O}(\epsilon^{1/4})$ in \cite{3}. Taking $a=0.998$ as a limit, one
can obtain the maximal CM energy per unit mass $19.20$, which is a
finite value. Meanwhile, Lake showed that the CM energy of the
collision at the inner horizon of the black hole is generically
divergent \cite{4} and the colliding particles can have arbitrary angular
momentum per unit rest mass that could fall into the black hole. But
he claimed soon after \cite{4} that the collision at the inner horizon actually
could not take place \cite{5}, which leads to no divergence of the CM energy. On the other hand, Grib and Pavlov suggested that the CM
energy can be unlimited in the case of multiple scattering
\cite{Grib,Grib1,Grib2,Grib3}. The universal property of
acceleration of particles for black holes was investigated in
\cite{Zaslavskii,Zaslavskii1,g}. In \cite{wei,wei1}, the property of
the CM energy for two colliding particles in the background of a
charged spinning black hole was discussed. One of the important
results of \cite{wei,wei1} is that the CM energy can still be
unlimited despite the deviation of the spin from its maximal value.
The BSW approach was then applied to the collision of particles
plunging from the innermost stable circular orbit and last stable
orbit near the horizon in \cite{xiao,xiao1}. There are also some
investigations on naked
singularities~\cite{Nake1,Nake2,Nake3,Nake4} and (anti-) de Sitter
backgrounds~\cite{ads}.

In this paper, we will investigate the property of the
collision of particles in the background of Einstein-Maxwell-dilaton
gravity. Firstly we calculate geodesic equations in Kaluza-Klein
spacetime in Einstein-Maxwell-dilaton theory. Then, we study the CM
energy for collisions taking place at the horizon of three cases
of Kaluza-Klein black holes. We find the CM energy is
unlimited for a pair of point particles colliding at the horizon,
with some fine tuning for the charged nonrotating case and the
extremal rotating case. The result of the near-extremal case
shows that the CM energy is in fact limited because the critical angular
momentum is too large for the geodesics of particles to reach the
horizon. Then we obtain the numerical result of the
maximal CM energy per unit mass for some different value of
$\epsilon=1-a$ in the case of a near-extremal Kaluza-Klein black hole, due to the difficulty in finding the exact result. Lastly, the Kaluza-Klein naked singularity is also considered in the present work. We find that the CM energy of collision between two particles can be arbitrarily high in this case.

This paper is organized as follows. In Section 2, we give a brief
review of four-dimensional Kaluza-Klein black holes and obtain the
geodesic equations in the background of Kaluza-Klein spacetime. In Section 3, we study the CM energy for collisions
taking place at the horizon for different cases of
Kaluza-Klein black holes and in the naked singularity case. The last section is devoted to discussion and conclusions.

\section{Geodesic equations in Kaluza-Klein spacetime}

We begin with a brief review of the Kaluza-Klein black hole.
It is derived by a dimensional
reduction of the boosted five-dimensional Kerr solution to four
dimensions. It is an exact solution of Einstein-Maxwell-Dilaton
theory. The metric is explicitly given by \cite{kk}
\begin{eqnarray}\label{metric}
ds^2&=&-\frac{1-Z}{B}dt^2-\frac{2aZ\sin^2\theta}{B\sqrt{1-\nu^2}}dtd\varphi+\frac{B\Sigma}{\Delta}dr^2+B\Sigma
d\theta^2\nonumber\\&&+\left[B(r^2+a^2)+a^2\sin^2\theta\frac{Z}{B}\right]\sin^2\theta
d\varphi^2 \;\;,
\end{eqnarray}
where
\begin{eqnarray}
&&Z=\frac{2\mu r}{\Sigma}\;\;,\nonumber\\
&&B=\sqrt{1+\frac{\nu^2Z}{1-\nu^2}}\;\;,\nonumber\\
&&\Sigma=r^2+a^2\cos^2\theta\;\;,\nonumber\\
&&\Delta=r^2-2\mu r+a^2.
\end{eqnarray}
The gauge potential is
\begin{eqnarray}
A=\frac{\nu}{2(1-\nu^2)}\frac{Z}{B^2}dt-\frac{a\nu\sin^2\theta}{2\sqrt{1-\nu^2}}\frac{Z}{B^2}d\varphi\;\;.
\end{eqnarray}
The physical mass $M$, the charge $Q$, and the angular momentum $J$
are expressed in terms of the boost parameter $\nu$, the mass parameter $\mu$,
and the specific angular momentum $a$ as
\begin{eqnarray}
M&=&\mu\left[1+\frac{\nu^2}{2(1-\nu^2)}\right]\;\;,\nonumber\\
Q&=&\frac{\mu\nu}{1-\nu^2}\;\;,\nonumber\\
J&=&\frac{\mu a}{\sqrt{1-\nu^2}}\;\;.
\end{eqnarray}
The outer and inner horizons are respectively defined at
\begin{eqnarray}
r_{\pm}=\mu\pm\sqrt{\mu^2-a^2}\;\;.
\end{eqnarray}
Thus, $\mu=a$ corresponds to the extremal black hole with one
degenerate horizon. The components of the inverse metric are
\begin{eqnarray}
g^{tt}&=&-\frac{B(r^2+a^2)+a^2\sin^2\theta\frac{Z}{B}}{\Delta}\;\;,\nonumber\\
g^{rr}&=&\frac{\Delta}{B\Sigma}\;\;,g^{\theta\theta}=\frac{1}{B\Sigma}\;\;,\nonumber\\
g^{\varphi\varphi}&=&\frac{1-Z}{B\Delta\sin^2\theta}\;\;,g^{t\varphi}=-\frac{aZ}{B\Delta\sqrt{1-\nu^2}}\;\;.
\end{eqnarray}

Because there are two Killing vectors $(\frac{\partial}{\partial
t})^{\mu}$ and $(\frac{\partial}{\partial \varphi})^{\mu}$, we have
two conserved quantities along a geodesic motion for a test particle
with charge $e$ as follows
\begin{eqnarray}
E&=&-g_{\mu\sigma}(\frac{\partial}{\partial t})^{\mu}[u^{\sigma}+eA^{\sigma}]=-(g_{tt}\dot{t}+g_{t\varphi}\dot{\varphi}-eA_{t})\;\;,\\
L&=&g_{\mu\sigma}(\frac{\partial}{\partial
\varphi})^{\mu}[u^{\sigma}+eA^{\sigma}]=g_{t\varphi}\dot{t}+g_{\varphi\varphi}\dot{\varphi}-eA_{\varphi}\;\;,
\end{eqnarray}
where $E$ and $L$ correspond to the constant energy and angular
momentum along a geodesic motion, respectively. It is easy to solve
the above equations for $\dot{t}$ and $\dot{\varphi}$ as
\begin{eqnarray}\label{2b}
\dot{t}&=&\frac{B(r^2+a^2)+a^2\frac{Z}{B}\sin^2\theta}{\Delta}(E-eA_{t})-\frac{aZ}{B\Delta\sqrt{1-\nu^2}}(L+eA_{\varphi})\nonumber\;\;,\\
\dot{\varphi}&=&\frac{aZ}{B\Delta\sqrt{1-\nu^2}}(E-eA_{t})+\frac{1-Z}{B\Delta\sin^2\theta}(L+eA_{\varphi})\;\;.
\end{eqnarray}
Substituting these solutions into the normalization condition
$g_{\mu\nu}u^{\mu}u^{\nu}=-1$ on the equatorial plane,
$\theta=\frac{\pi}{2}$ and $\dot{\theta}=0$, one will arrive at
\begin{eqnarray}\label{22}
\dot{r}=\frac{\Delta}{B\Sigma}R(r)\;\;,
\end{eqnarray}
where
\begin{eqnarray}
\frac{\Delta}{B\Sigma}R^2(r)&=&\frac{B(r^2+a^2)+a^2\frac{Z}{B}}{\Delta}(E-eA_{t})^2-2\frac{aZ}{B\Delta\sqrt{1-\nu^2}}(E-eA_{t})(L+eA_{\varphi})\nonumber\\
&&-\frac{1-Z}{B\Delta}(L+eA_{\varphi})^2-1\;\;.
\end{eqnarray}
Now we have solved the geodesic equations on the equatorial plane
in Kaluza-Klein spacetime. In the next section, we will turn to the
CM energy for particles colliding in this background.

\section{Center-of-mass energy for collisions in
Kaluza-Klein spacetime}

The energy in the center-of-mass frame for a pair of point particles
colliding is computed by the formula \cite{1}
\begin{eqnarray}\label{2c}
E_{CM}=\sqrt{2}m_0\sqrt{1-g_{\mu\nu}u_1^\mu u_2^\nu}\;\;,
\end{eqnarray}
where $u_1^\mu$ and $u_2^\nu$ are the 4-velocities of the two
particles. For the case that the particles begin at rest at infinity
and the collision energy comes solely from gravitational
acceleration, the particles follow geodesics
with energy $E\geq1$. Consider two particles coming from
infinity with $E_1=E_2=1$ and approaching the black hole with
different angular momenta $L_1$ and $L_2$. Taking into account the
metric of the Kaluza-Klein black hole (\ref{metric}) on the equatorial
plane, we obtain the CM energy for collision with the help of
(\ref{2b}) (\ref{22}) and (\ref{2c})

\begin{eqnarray}\label{4}
\frac{E_{CM}^2}{2m_0^2}=1+\frac{K}{B\Delta}\;\;,
\end{eqnarray}
where
\begin{eqnarray}\label{5}
K&=&[B^2(r^2+a^2)+a^2Z](1-e_1A_{t})(1-e_2A_{t})+(Z-1)(L_1+e_1A_{\varphi})(L_2+e_2A_{\varphi})\nonumber\\
&&-\frac{aZ}{\sqrt{1-\nu^2}}[(1-e_1A_{t})(L_2+e_2A_{\varphi})+(1-e_2A_{t})(L_1+e_1A_{\varphi})]\nonumber\\
&&-\bigg\{[B^2(r^2+a^2)+a^2Z](1-e_1A_{t})^2+(Z-1)(L_1+e_1A_{\varphi})^2\nonumber\\
&&-\frac{2aZ}{\sqrt{1-\nu^2}}(1-e_1A_{t})(L_1+e_1A_{\varphi})-B\Delta\bigg\}^\frac{1}{2}\nonumber\\
&&\times\bigg\{[B^2(r^2+a^2)+a^2Z](1-e_2A_{t})^2+(Z-1)(L_2+e_2A_{\varphi})^2\nonumber\\
&&-\frac{2aZ}{\sqrt{1-\nu^2}}(1-e_2A_{t})(L_2+e_2A_{\varphi})-B\Delta\bigg\}^\frac{1}{2}\;\;.
\end{eqnarray}
We have obtained the CM energy of two colliding particles in Kaluza-Klein spacetime. Now we are ready to investigate the CM energy for
different cases of Kaluza-Klein black holes and for the naked singularity case.

\subsection{Charged nonrotating Kaluza-Klein black hole }

The acceleration of particles by a Reissner-Nordstr\"{o}m black hole has
been discussed in \cite{Zaslavskii1}. Charged nonrotating Kaluza-Klein spacetime is
very different from the Reissner-Nordstr\"{o}m case because there is
only one event horizon, which leads to the absence of an extremal
nonrotating Kaluza-Klein black hole. The metric of the charged
nonrotating Kaluza-Klein spacetime is
\begin{eqnarray}\label{3a}
ds^2&=&-\frac{\Delta}{r^2B}dt^2+\frac{r^2B}{\Delta}dr^2+Br^2d\theta^2+Br^2\sin^2\theta
d\varphi^2 \;\;,
\end{eqnarray}
where
\begin{eqnarray}
&&B=\sqrt{1+\frac{\nu^2Z}{1-\nu^2}}\;\;,\nonumber\\
&&Z=\frac{2\mu}{r}\;\;,\nonumber\\
&&\Delta=r^2-2\mu r\;\;.
\end{eqnarray}
The gauge potential is given by
\begin{eqnarray}
A=\frac{\nu}{2(1-\nu^2)}\frac{Z}{B^2}dt\;\;.
\end{eqnarray}
The horizon lies at $r_h=2\mu$. According to Eq.~(\ref{4}) and
Eq.~(\ref{5}), we can obtain the CM energy of two radial motion
particles colliding in charged nonrotating Kaluza-Klein spacetime as
\begin{eqnarray}
\frac{E_{CM}^2}{2m_0^2}=1+\frac{K_1}{B\Delta}\;\;,
\end{eqnarray}
where
\begin{eqnarray}
K_1&=&B^2r^2(1-e_1A_{t})(1-e_2A_{t})-\sqrt{[B^2r^2(1-e_1A_{t})^2-B\Delta][B^2r^2(1-e_2A_{t})^2-B\Delta]}\;\;.
\end{eqnarray}
It appears that $E_{CM}^2$ diverges at $r=r_h$, but this is not true
because, although not totally obvious, the numerator vanishes at
that point as well. After some calculations, the CM energy is
\begin{eqnarray}
\frac{E_{CM}^2}{2m_0^2}&=&1+\frac{1}{2}\left(\frac{1-e_2\frac{\nu}{2}}{1-e_1\frac{\nu}{2}}+\frac{1-e_1\frac{\nu}{2}}{1-e_2\frac{\nu}{2}}\right)\;\;.
\end{eqnarray}
If one of the particles participating in the collision has the
critical charge $e=\frac{2}{\nu}$, the CM energy will blow up at the
horizon. Thus we have shown that non-extremal black holes could also
serve as particle accelerators and provide a visible probe
of Planck-scale physics.

\subsection{Extremal charged rotating Kaluza-Klein black hole }
In the case $a=\mu$, which corresponds to the extremal Kaluza-Klein
black hole, we obtain the form of the CM energy of two uncharged
particles colliding at the degenerate horizon, after some tedious
calculations, as:
\begin{eqnarray}\label{3c}
E_{CM}^{KK}(r\rightarrow
r_+)&=&\sqrt{2}m_0\bigg[1+\frac{(1+\nu^2)(L_1-L_2)^2}{2\sqrt{1-\nu^4}(L_1-\frac{2\mu}{\sqrt{1-\nu^2}})(L_2-\frac{2\mu}{\sqrt{1-\nu^2}})}\nonumber\\
&&+\frac{1}{2}\left(\frac{L_1-\frac{2\mu}{\sqrt{1-\nu^2}}}{L_2-\frac{2\mu}{\sqrt{1-\nu^2}}}+\frac{L_2-\frac{2\mu}{\sqrt{1-\nu^2}}}{L_1-\frac{2\mu}{\sqrt{1-\nu^2}}}\right)\bigg]^{\frac{1}{2}}\;\;,
\end{eqnarray}

Clearly, when $L_1$ or $L_2$ takes the critical angular momentum
$L_c=\frac{2\mu}{\sqrt{1-\nu^2}}$, the CM energy $E^{KK}_{CM}$ will
be unlimited, which means that the particles can collide with
arbitrarily high CM energy at the horizon. We expect that, in the
case $\nu=0$, the CM energy (\ref{3c}) in the background of a
Kaluza-Klein black hole should reduce to the one in the background
of a Kerr black hole. After some calculations, we find that the CM
energy is exactly consistent with that of \cite{1} in the case
$\nu=0$.

%%%%%%%%%%%%%%%%%%%%%%%%%%%%%%%%%%%%%%%%%%%%%%%%%%%%%%%%%%%%%%%%%%%%%%%%%%%%%
\begin{figure*}
\centerline{\subfigure[]{
\includegraphics[width=8cm,height=6cm]{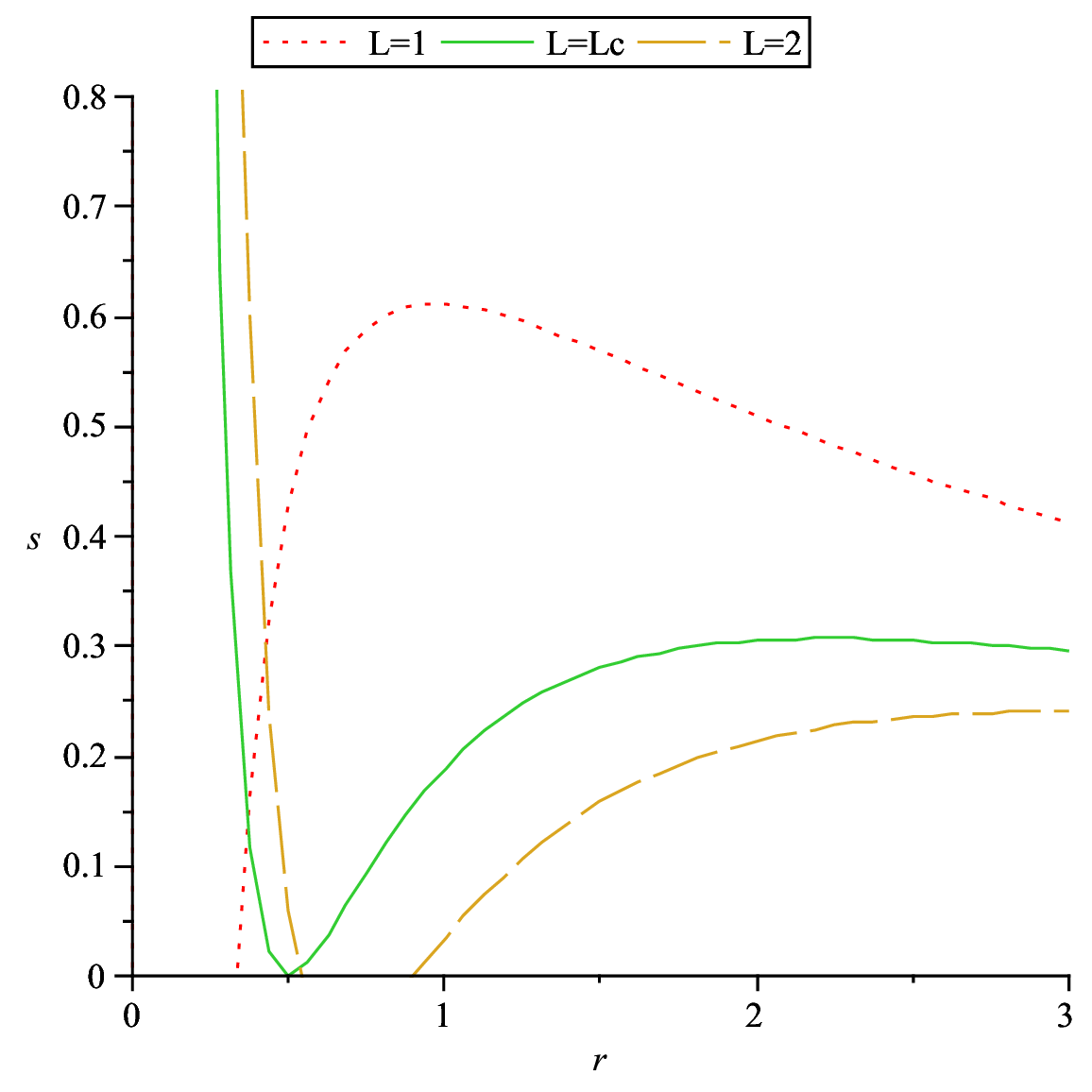}}
\subfigure[]{
\includegraphics[width=8cm,height=6cm]{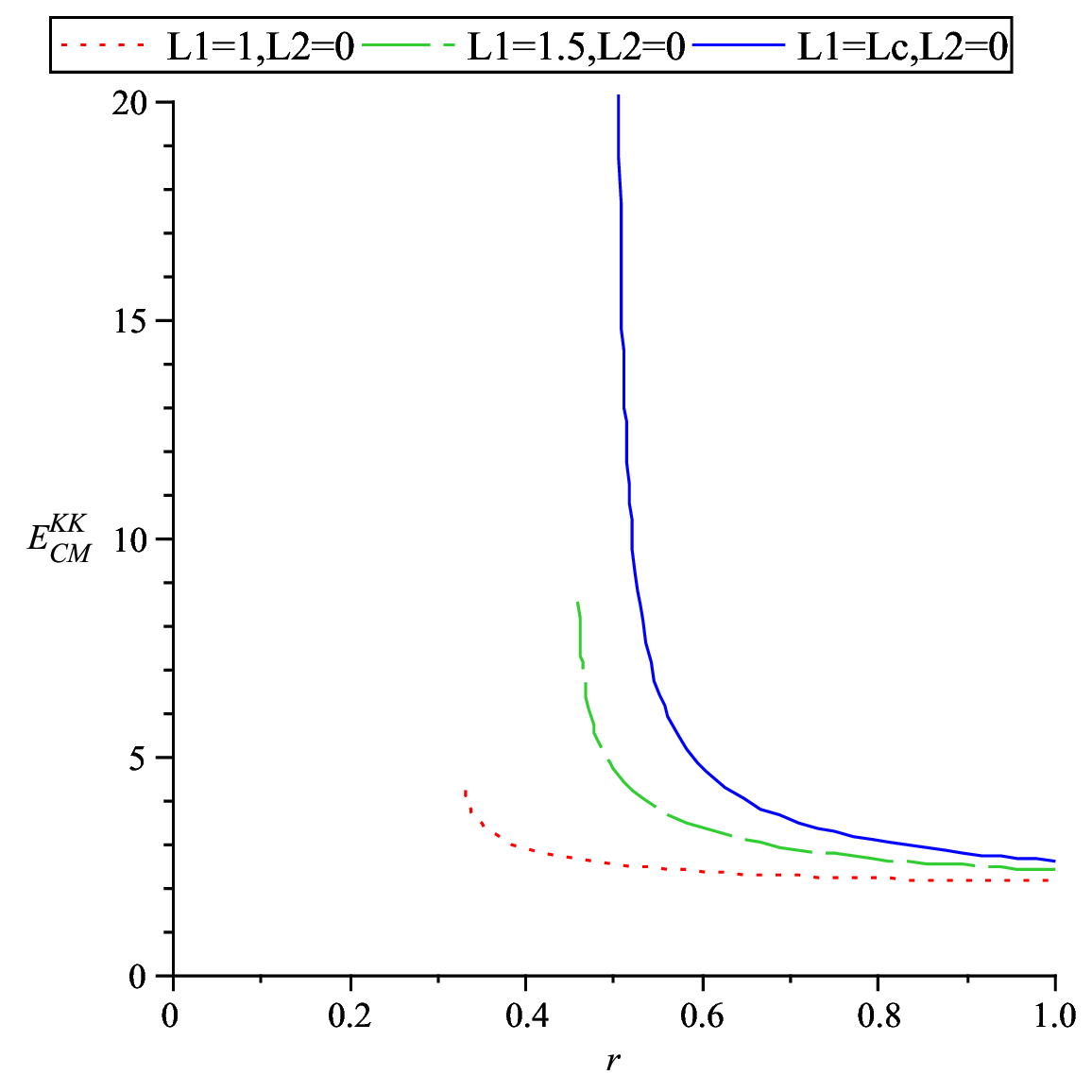}}}
\caption{For an extremal Kaluza-Klein Black hole with
$J=\frac{\sqrt{3}}{4}$ and $M=1$, (a) the variation of $s=\dot{r}^2$
with radius for three different values of angular momentum, and (b) the
variation of $E^{KK}_{CM}$ with radius for three combinations of
$L_1$ and $L_2$. For $L_1=L_c$, $E^{KK}_{CM}$ blows up at the
horizon.} \label{1}
\end{figure*}
%%%%%%%%%%%%%%%%%%%%%%%%%%%%%%%%%%%%%%%%%%%%%%%%%%%%%%%%%%%%%%%%%%%%%%%%%%%%%%%%

%%%%%%%%%%%%%%%%%%%%%%%%%%%%%%%%%%%%%%%%%%%%%%%%%%%%%%%%%%%%%%%%%%%%%%%%%%%%%
\begin{figure*}
\centerline{\subfigure[]{
\includegraphics[width=8cm,height=6cm]{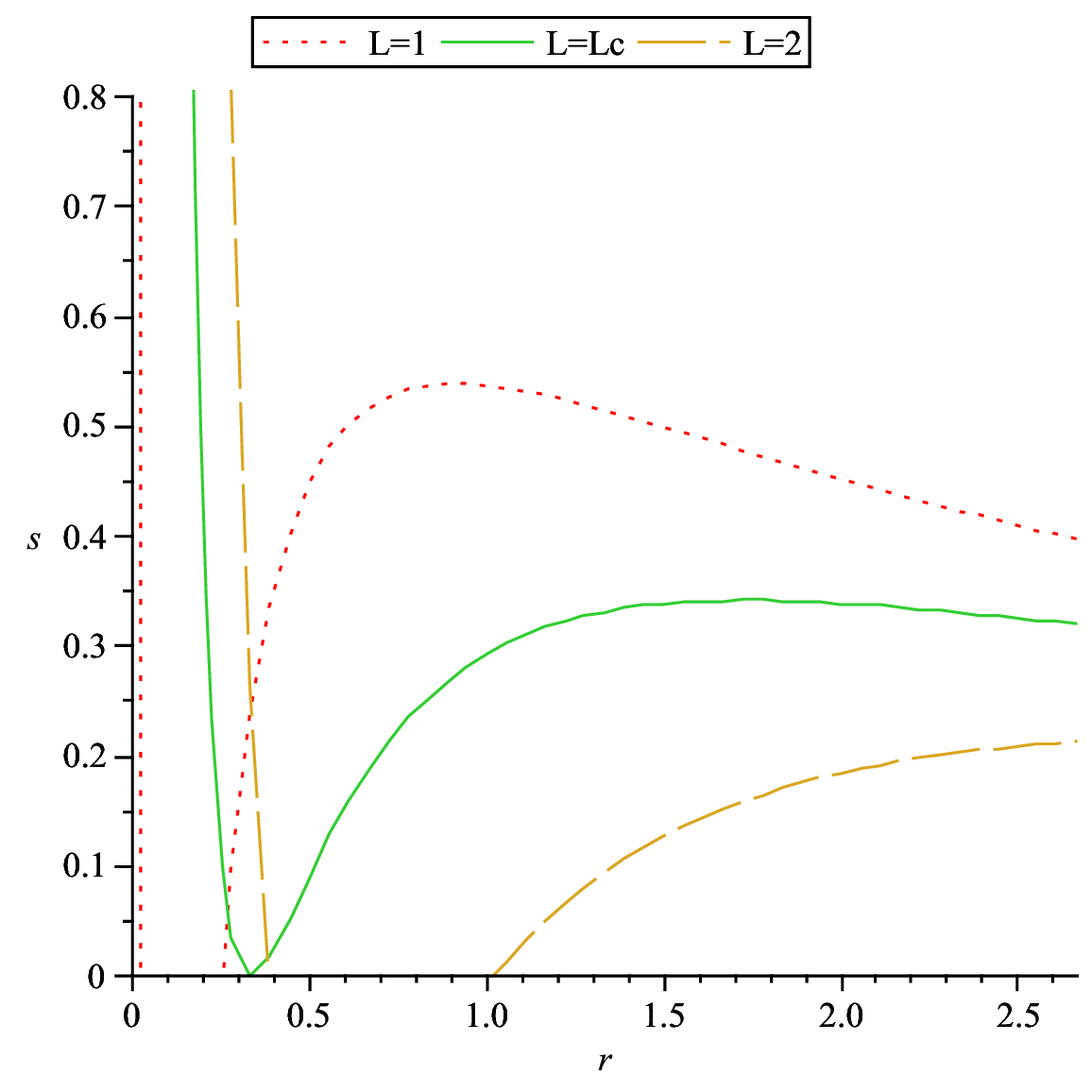}}
\subfigure[]{
\includegraphics[width=8cm,height=6cm]{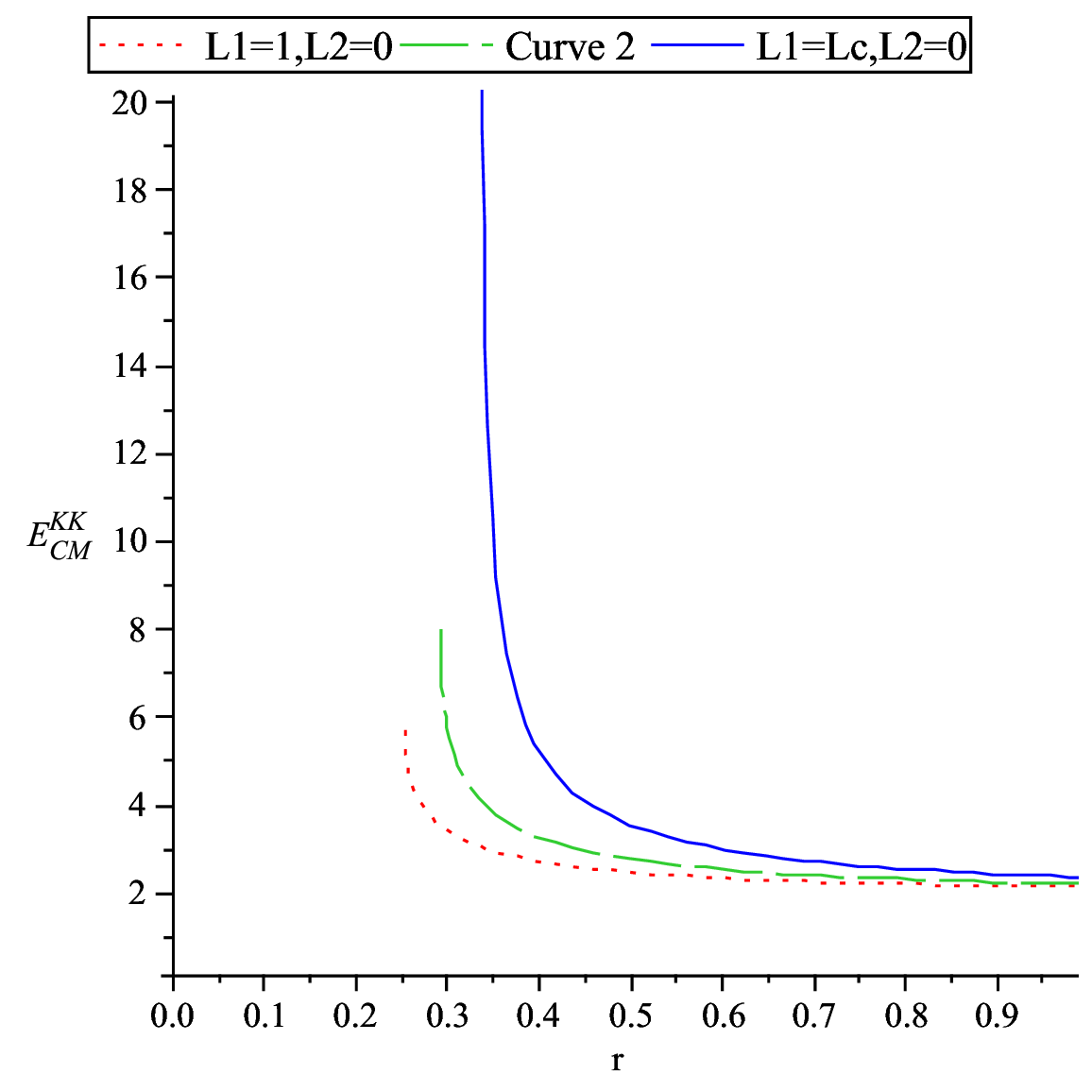}}}
\caption{For an extremal Kaluza-Klein Black hole with
$J=\frac{\sqrt{5}}{9}$ and $M=1$, (a) the variation of $s=\dot{r}^2$
with radius for three different values of angular momentum, and (b) the
variation of $E=E^{KK}_{CM}$ with radius for three combinations of
$L_1$ and $L_2$. For $L_1=L_c$, $E^{KK}_{CM}$ blows up at the
horizon. } \label{2}
\end{figure*}
%%%%%%%%%%%%%%%%%%%%%%%%%%%%%%%%%%%%%%%%%%%%%%%%%%%%%%%%%%%%%%%%%%%%%%%%%%%%%%%%

We plot $\dot{r}^2$ and $E^{KK}_{CM}$ in Fig.~\ref{1} and
Fig.~\ref{2}, from which we can see that there exists a critical
angular momentum $L_c=\frac{2\mu}{\sqrt{1-\nu^2}}$ for the geodesics
of particle to reach the horizon. If $L>L_c$, the geodesics never
reach the horizon. On the other hand, if the angular momentum is too
small, the particle will fall into the black hole and the CM energy
for the collision is limited. However, when $L_1$ or $L_2$ takes the
angular momentum $L=\frac{2\mu}{\sqrt{1-\nu^2}}$, the CM energy is
unlimited. As a result, it may provide a
unique probe for Planck-scale physics.

\subsection{Near-extremal charged rotating Kaluza-Klein black hole }
For the near-extremal case, we also obtain the CM energy of two uncharged particles colliding at the
outer horizon:
\begin{eqnarray}\label{3d}
E_{CM}^{KK}(r\rightarrow
r_+)&=&\sqrt{2}m_0\bigg[1+\frac{(\frac{r_+^2}{a^2}+\nu^2)(L_1-L_2)^2}{2B(r_+)(1-\nu^2)(L_1-\frac{r_+^2+a^2}{a\sqrt{1-\nu^2}})(L_2-\frac{r_+^2+a^2}{a\sqrt{1-\nu^2}})}\nonumber\\
&&+\frac{1}{2}\left(\frac{L_1-\frac{r_+^2+a^2}{a\sqrt{1-\nu^2}}}{L_2-\frac{r_+^2+a^2}{a\sqrt{1-\nu^2}}}+\frac{L_2-\frac{r_+^2+a^2}{a\sqrt{1-\nu^2}}}{L_1-\frac{r_+^2+a^2}{a\sqrt{1-\nu^2}}}\right)\bigg]^{\frac{1}{2}}\;\;.
\end{eqnarray}

\begin{table}[h]
\begin{center}
\caption{The CM energy per unit rest mass
$\frac{E_{\text{cm}}}{m_{0}}$ for a KK black hole with spin
$a=1-\epsilon$ and $L_{1}=L_{\text{max}}$,
$L_{2}=0$.}\label{TableCMenergy}
\begin{tabular}{c c c c c c }
\hline
\hline
& $\epsilon$=0.1 & $\epsilon$=0.01 & $\epsilon$=0.001 & $\epsilon$=0.0001  & $\epsilon$=0.00001\\
\hline
$\nu$=0\;\;\;\;& 4.21767 \;\;& 7.10481 \;\;& 12.43999 \;\;& 22.01962 \;\;& 39.10101 \\
$\nu$=0.1 \;\;& 4.21931 \;\;& 7.10426 \;\;& 12.43576 \;\;& 22.00991 \;\;& 39.08243 \\
$\nu$=0.2 \;\;& 4.22482 \;\;& 7.10430 \;\;& 12.42667 \;\;& 21.98763 \;\;& 39.03917 \\
$\nu$=0.3 \;\;& 4.23623 \;\;& 7.11010 \;\;& 12.42368 \;\;& 21.97348 \;\;& 39.00878 \\
$\nu$=0.4 \;\;& 4.25740 \;\;& 7.13106 \;\;& 12.44599 \;\;& 22.00338 \;\;& 39.05618 \\
$\nu$=0.5 \;\;& 4.29529 \;\;& 7.18270 \;\;& 12.52408 \;\;& 22.13337 \;\;& 39.28214 \\
$\nu$=0.6 \;\;& 4.36278 \;\;& 7.29163 \;\;& 12.70810 \;\;& 22.45441 \;\;& 39.84939 \\
$\nu$=0.7 \;\;& 4.48641 \;\;& 7.50928 \;\;& 13.09206 \;\;& 23.13507 \;\;& 41.05854 \\
$\nu$=0.8 \;\;& 4.73260 \;\;& 7.95838 \;\;& 13.89599 \;\;& 24.56740 \;\;& 43.60709 \\
$\nu$=0.9 \;\;& 5.34624 \;\;& 9.07637 \;\;& 15.89588 \;\;& 28.12961 \;\;& 49.94483 \\
\hline
\end{tabular}
\end{center}
\end{table}

Naively, the CM energy $E^{KK}_{CM}$
will be divergent when $L_1$ or $L_2$ takes the critical angular momentum
\be
L_c=\frac{r_+^2+a^2}{a\sqrt{1-\nu^2}}.
\ee
However, a careful analysis shows that the critical angular momentum $L_c$ is too large for the geodesics of the particle to reach the
horizon. That is, a freely falling particle with this critical angular momentum will be reflected before it reaches the horizon. The turning point of an initially ingoing particle is located at the larger root of the equation
\be
V_{eff}=0,
\ee
where $V_{eff}$ is the effective potential for the radial motion. For an uncharged particle coming from infinity with $E=1$, it is given by
\be
V_{eff}=\frac{1}{B^2 r^4}\left[(a L-\frac{r^2+a^2}{\sqrt{1-\nu^2}})^2-\Delta(L-\frac{a}{\sqrt{1-\nu^2}})^2-\Delta\frac{r^2\nu^2}{1-\nu^2}-\Delta B r^2\right],
\ee
where $\Delta=r^2-2\mu r+a^2$ and $B=\sqrt{1+\frac{2\mu \nu^2}{r(1-\nu^2)}}$. We find that $V_{eff}=0$ at the horizon when the freely falling particle has the critical angular momentum $L_c$. However, one can derive at the horizon that
\be
\frac{d\,V_{eff}}{d\,r}\mid_{_{r=r_+}}=-\frac{\Delta'}{B^2r_+^4}\left[(L_c-\frac{a}{\sqrt{1-\nu^2}})^2+\frac{r_+^2\nu^2}{1-\nu^2}+ B r_+^2\right],
\ee
where $\Delta'=\frac{d\,\Delta}{d\,r}\mid_{_{r=r_+}}$ and $\Delta\mid_{_{r=r_+}}=0$ has been used. Since $r_+$ is the larger root of $\Delta=0$, $\Delta'=r_+-r_->0$. Given the fact that the boost parameter
$\nu^2<1$, hence $\frac{d\,V_{eff}}{d\,r}\mid_{r=r_+}<0$. That means the effective potential will decrease when $r$ grows from $r_+$. Thus there will be at least one root located at $r_+<r<\infty$. Consequently, the freely falling particle with critical angular momentum $L_c$ will be reflected before reaching the horizon. Hence the arbitrarily high CM energy will not be achieved in the case of collision of two freely falling particles from infinity. For particles which can reach the horizon, their angular momentum should be smaller than the angular momentum of a circular orbit particle. We will refer to the angular momentum of a circular orbit particle as the maximal angular momentum $L_{max}$. The angular momentum of a circular orbit particle and the radius of the circular orbit are defined implicitly from the solution of the following equations
\be\label{Lmax}
V_{eff}=\frac{d\,V_{eff}}{d\,r}=0.
\ee
Though an analytical definition of $L_{max}$ is very hard to get from Eq.~(\ref{Lmax}) due to the presence of $r$ in the expression of $B$, one can always solve it numerically when other parameters $\mu,\nu,a$ are specified. Taking the unit $\mu=1$ and one of the particles falling with $L_{max}$, while the other particle falls without orbital angular momentum, we list the numerical results of
 CM energy per unit mass in Table \ref{TableCMenergy} for different small parameter
$\epsilon=1-a$ and different values of $\nu$. The results show that the CM energy is in fact limited and grows slowly as the black hole spin
approaches its maximal value.

\subsection{Naked singularity case}
Lastly, we will consider the naked singularity case. If
we take a near-extremal Kaluza-Klein naked singularity, we expect that
it is possible for an ingoing and an outgoing
particle to collide, just like the case of the Kerr naked singularity \cite{Nake3}. Then the CM energy for
collision in the Kaluza-Klein naked singularity is
\begin{eqnarray}\label{v}
\frac{E_{CM}^2}{2m_0^2}=1+\frac{\mathcal {K}}{B\Delta}\;\;,
\end{eqnarray}
where
\begin{eqnarray}
\mathcal {K}&=&[B^2(r^2+a^2)+a^2Z](1-e_1A_{t})(1-e_2A_{t})+(Z-1)(L_1+e_1A_{\varphi})(L_2+e_2A_{\varphi})\nonumber\\
&&-\frac{aZ}{\sqrt{1-\nu^2}}[(1-e_1A_{t})(L_2+e_2A_{\varphi})+(1-e_2A_{t})(L_1+e_1A_{\varphi})]\nonumber\\
&&+\bigg\{[B^2(r^2+a^2)+a^2Z](1-e_1A_{t})^2+(Z-1)(L_1+e_1A_{\varphi})^2\nonumber\\
&&-\frac{2aZ}{\sqrt{1-\nu^2}}(1-e_1A_{t})(L_1+e_1A_{\varphi})-B\Delta\bigg\}^\frac{1}{2}\nonumber\\
&&\times\bigg\{[B^2(r^2+a^2)+a^2Z](1-e_2A_{t})^2+(Z-1)(L_2+e_2A_{\varphi})^2\nonumber\\
&&-\frac{2aZ}{\sqrt{1-\nu^2}}(1-e_2A_{t})(L_2+e_2A_{\varphi})-B\Delta\bigg\}^\frac{1}{2}\;\;.
\end{eqnarray}
Let us work in the unit $\mu=1$. If the collision happens to take place at $r=1$, the CM energy of collision between two particles can be very high in the limit that the deviation of an extremal Kaluza-Klein naked singularity is very small ($\Delta\rightarrow0$). Essentially, the divergence of the CM energy comes from the extremality of the naked singularity. Nonetheless, the range of allowed angular momentum of a freely falling particle is not arbitrary, to guarantee that it can turn back at some point to become an outgoing particle. The turning point of an initially ingoing particle is located at the larger root of the equation
\be\label{eff}
V_{eff}=0,
\ee
where $V_{eff}$ is the effective potential for the radial motion. For an uncharged particle coming from infinity with $E=1$, it is given by
\begin{align}
V_{eff}=-\big[1+\frac{a^2}{r^2}+\frac{2a^2}{r^3B^2}-\frac{4a L}{r^3B^2\sqrt{1-\nu^2}}-\frac{(r-2)L^2}{r^3B^2}-\frac{r^2-2 r+a^2}{Br^2}\big],
\end{align}
where $B=\sqrt{1+\frac{2 \nu^2}{r(1-\nu^2)}}$. For the case without a real root of Eq.~(\ref{eff}), the particle will hit the singularity eventually. To have a collision of the two particles at $r=1$, one just needs to solve  the range of allowed angular momentum from Eq.~(\ref{eff}) to guarantee that the following condition is satisfied: one of the particles should have a turning point at $r_t<1$ and the other particle should not turn back before reaching $r=1$. Though the analytical solution to Eq.~(\ref{eff}) is extremely hard, if not impossible, to get, one can always check the existence of a solution numerically when all the parameters are specified.

\section{Discussion and Conclusions}

In this paper, we have investigated the CM energy for
two colliding particles in Kaluza-Klein spacetime. The Kaluza-Klein black hole is an exact
solution in Einstein-Maxwell-dilaton theory in four-dimensional
spacetime. When the charge $Q$ vanishes, it just describes the Kerr
black hole. Hence there is a restriction that our result should not be
in contradiction with that of the Kerr black hole when $Q=0$, which has been
proved by the calculations. Our results show that an extremal
Kaluza-Klein black hole can serve as a particle accelerator with
arbitrarily high CM energy when one of the colliding particles has
the fine-tuned angular momentum $L=\frac{2\mu}{\sqrt{1-\nu^2}}$. For the
near-extremal case, in terms of the small parameter $\epsilon=1-a$,
we also obtain the numerical result of the maximal CM energy per
unit mass for some different values of $\epsilon$ and $\nu$. Our
near-extremal result shows that the CM energy will not be so high,
even in the very near-extremal case. On the other hand, with
the vanished angular momentum $J$, the Kaluza-Klein black hole does
not reduce to the Reissner-Nordstr\"{o}m black hole. Our result in
the nonrotating case shows that the CM energy of two charged
colliding particles could also blow up, thus a non-extremal black
hole could also provide
a visible probe of Planck-scale physics. In both cases where arbitrarily high CM energy can be reached, a very specific angular momentum of the colliding particle needs to be chosen. Hence one must fine-tune the angular momentum of the ingoing particle. To overcome this issue, we also studied the collision of particles in the Kaluza-Klein naked singularity case. Our results show that the range of allowed angular momenta of freely falling particles to reach arbitrarily high CM energy will be much larger than a single fine-tuned value.

However, our calculations were performed without considering the back
reaction effect of the accelerated particle pair on the background
geometry of the Kaluza-Klein black hole. It should be pointed out that particles can
be accelerated to arbitrarily high CM energy. Hence the background
geometry may be destroyed and the back reaction effect should not be
ignored. On the other hand, high energy concentrated at small scale
will lead to gravitational collapse. So the Planck-scale physics
induced by the collision of a particle pair with arbitrarily high CM
energy is protected by the event horizon formed due to gravitational
collapse, and cannot be observed externally.
Hence, it would definitely be  interesting and meaningful to explore the field theory interpretation of this classical
effect in the future.

\begin{acknowledgments}
\noindent This work is supported in part by NSFC Grant No. 11575202 and No. 11205048, by the Foundation for Young Key Teacher of Henan Normal University and by the Cuiying Programme of Lanzhou University (225000-582404) and the Fundamental Research Fund for
Physics and Mathematic of Lanzhou University (LZULL200911). P.J.M. is grateful to Xiaolei Sun and Shi-Xiong
Song for valuable information and helpful discussions.
\end{acknowledgments}

\end{CJK*}
\end{document}